\documentclass[10pt]{article}
\usepackage{graphicx}
\usepackage{multirow}

\def\Title#1{\begin{center} {\Large #1 } \end{center}}
\def\Author#1{\begin{center}{ \sc #1} \end{center}}
\def\Address#1{\begin{center}{ \it #1} \end{center}}

\newcommand\pubblock{\rightline{\begin{tabular}{l} Proceedings of the Second Annual LHCP\\ \pubnumber\\
         \pubdate  \end{tabular}}}

\newenvironment{Abstract}{\begin{quotation} \begin{center} 
             \large ABSTRACT \end{center}\bigskip 
      \begin{center}\begin{large}}{\end{large}\end{center} \end{quotation}}

\newenvironment{Presented}{\begin{quotation} \begin{center} 
             PRESENTED AT\end{center}\bigskip 
      \begin{center}\begin{large}}{\end{large}\end{center} \end{quotation}}

\def\beq{\begin{equation}}
\def\eeq#1{\label{#1}\end{equation}}
\def\eeqn{\end{equation}}

\def\beqa{\begin{eqnarray}}
\def\eeqa#1{\label{#1}\end{eqnarray}}
\def\eeqan{\end{eqnarray}}

\let\bar=\overbar

\def\Dslash{\not{\hbox{\kern-4pt $D$}}}
\def\dslash{\not{\hbox{\kern-2pt $\del$}}}

\def\msb{{\bar{\ssstyle M \kern -1pt S}}}

\def\wwjj{$W^{\pm}W^{\pm}$jj}

\usepackage{color}

\newcommand\pubnumber{ ATL-PHYS-PROC-2014-115 }

\newcommand\pubdate{\today}

\def\affiliation{
Lawrence Berkeley National Laboratory \\
Berkeley, CA 94720, U.S.A., \\
On behalf of the ATLAS Collaboration
}

\begin{document}

\large
\begin{titlepage}
\pubblock

\vfill
\Title{ Electroweak production of $Z$jj and $W^{\pm}W^{\pm}$jj states at the LHC }
\vfill

\Author{ Simone Pagan Griso  }
\Address{\affiliation}
\vfill
\begin{Abstract}

Measurements of fiducial cross sections for the electroweak production of two jets in association with a $Z$ boson and in association with a pair of same-electric-charge $W$ bosons are presented.
The measurements are performed using $20.3~$fb$^{-1}$ of proton-proton collision data collected at a center-of-mass energy of $\sqrt{s}=8$~TeV by the ATLAS experiment at the Large Hadron Collider.
The measured fiducial cross sections are in agreement with the Standard Model predictions.
Limits at 95\% confidence level are set on anomalous triple and quartic gauge couplings.
\end{Abstract}
\vfill

\begin{Presented}
The Second Annual Conference\\
 on Large Hadron Collider Physics \\
Columbia University, New York, U.S.A \\ 
June 2-7, 2014
\end{Presented}
\vfill
\end{titlepage}
\def\thefootnote{\fnsymbol{footnote}}
\setcounter{footnote}{0}
%

\normalsize 


\section{Introduction and measurements goals}

The electroweak production of two jets in association with a $Z/\gamma^\star$ boson ($Z$jj) or in association with a pair of same-electric-charge $W$ bosons (\wwjj), is a rare process. The expected cross section is in the range of few tens of $fb$ to one $fb$, after fiducial selections.
 Both processes have been recently observed by the ATLAS collaboration~\cite{Zjj,WWjj}, using $20.3~$fb$^{-1}$ of proton-proton collision data collected at a center-of-mass energy of $\sqrt{s}=8$~TeV by the ATLAS experiment~\cite{atlas} at the Large Hadron Collider (LHC). 

 The first evidence for electroweak \wwjj~processes represents a milestone in the investigation of the electroweak symmetry breaking, following a route complementary to the direct Higgs coupling measurements. In fact the scattering of two longitudinally polarized $W$ bosons violates unitarity without delicate cancellations from the Higgs boson contribution, and the measurement of \wwjj production cross section can constrain this contribution.
 The first observation of electroweak $Z$jj production also contributes to the investigation of electroweak symmetry breaking, being the first observation of a vector boson fusion process and being an irreducible background to many direct Higgs coupling measurements.
 In addition, these measurements can constrain anomalous triple and quartic gauge couplings which may arise from theories extending the Standard Model (SM).

\begin{figure}[htb]
\centering
\includegraphics[height=2in]{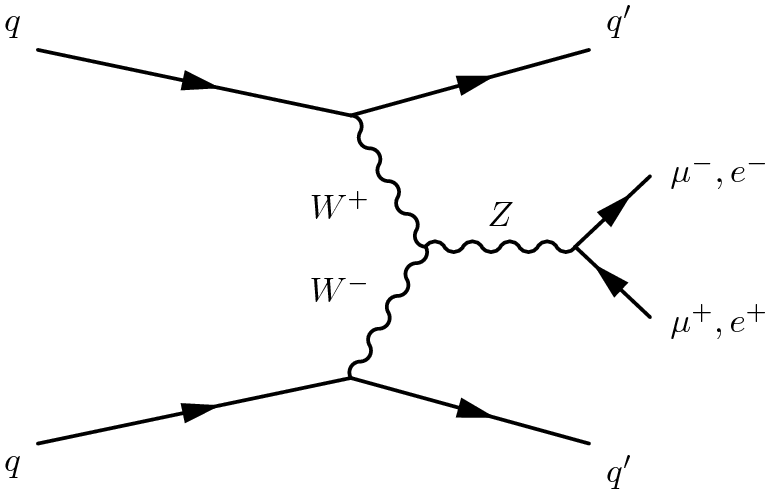}
\includegraphics[height=2in]{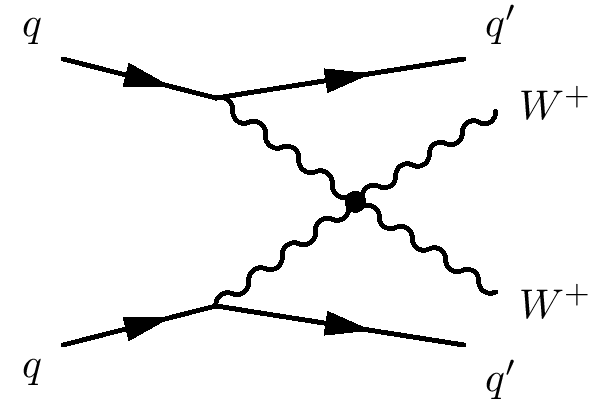}
\caption{ Representative Feynman diagrams for electroweak production of $Z$jj (left) and \wwjj~(right).}
\label{fig:FeynmanEWK}
\end{figure}

\begin{figure}[htb]
\centering
\includegraphics[height=2in]{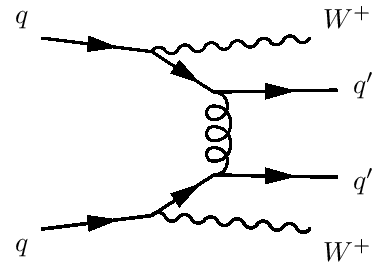}
\includegraphics[height=2in]{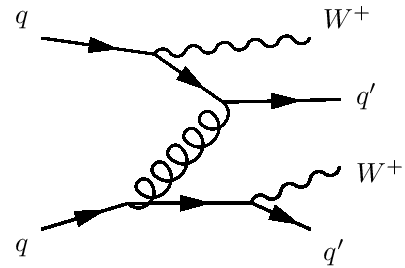}
\caption{ Representative Feynman diagrams for strong production of \wwjj~(right).}
\label{fig:FeynmanQCD}
\end{figure}

 The electroweak production of \wwjj~is defined using exclusively weak interactions at Born level (of order $\alpha_{EW}^{4}$ without considering the boson decay, where $\alpha_{EW}$ is the electroweak force coupling constant), as depicted in figure~\ref{fig:FeynmanEWK}~(right). Analogously, the electroweak production of $Z$jj is defined to be of order $\alpha_{EW}^{3}$ at Born level, as depicted in figure~\ref{fig:FeynmanEWK}~(left).
 The same final state can be produced using both the strong and electroweak interactions at Born level, and is referred to as strong production. Figure~\ref{fig:FeynmanQCD} shows an example of strong production, of order $\alpha_{s}^{2}\alpha_{EW}^{2}$~(where $\alpha_{s}$ is the strong force coupling constant), in the case of \wwjj.

In the case of $Z$jj production, the strong production dominate the production cross section for this final state, making extremely challenging the measurement and study of the purely electroweak component.
The case of \wwjj~production is quite peculiar among all similar di-boson final states, since the strong production cross section does not dominate the electroweak cross section; for this reason this channel is an ideal choice for isolating electroweak production in the scattering of vector bosons.

The electroweak production can be enhanced in appropriate regions of phase space. In particular a lack of additional jets between the two leading jets and a large invariant mass as well as rapidity separation of the leading jets can be expected in the case of electroweak production. This is due to the lack of color-connection between the two jets at Born level. 
The two measurements, make both use of such characteristics to isolate the electroweak component and suppress strong production.

The main goals of the two measurements presented in this contribution are very similar:
\begin{itemize}
\item Measure an inclusive (strong plus electroweak production) cross section and, in the case of $Z$jj where the statistics allows it, study the agreement of various kinematic distributions with simulation.
\item Isolate and measure the electroweak component in an appropriately designed phase space; in this case the strong production is considered as a background.
\item Set limits on anomalous triple or quartic gauge couplings; these results will not be discussed in detail in this contribution and we refer to~\cite{Zjj,WWjj} for the results and further details.
\end{itemize}

\section{Electroweak $Z$jj}
 Candidate events for the electroweak $Z$jj measurement are selected using a combination of single and di-lepton triggers. 
A summary of the selection criteria applied is presented in table~\ref{tab:ZjjRegions} for five different selections (regions).
The {\it baseline} region is constructed as an inclusive region to study $Z$ production in association with two jets. The {\it high-$p_{\rm T}$} and {\it high-mass} regions probe phase space particularly sensitive to the electroweak production and to contributions beyond the Standard Model. The {\it control} region is used to constraint the strong production modeling using data, while the {\it search} region is used to extract the electroweak component.

\begin{table}[t]
\begin{center}
\begin{tabular}{c|c|c|c|c|c}  
Object & {\it baseline} & {\it high-mass} & {\it search} & {\it control} & {\it high-$p_{T}$}\\
\hline
\hline
Leptons & \multicolumn{5}{|c}{$\vert\eta^{\ell}\vert < 2.47$, $p_{T}^{\ell} > 25$~GeV}\\
\hline
\multirow{2}{*}{Dilepton pair}& \multicolumn{5}{|c}{$81 < m(\ell\ell) < 101$~GeV}\\
                              & \multicolumn{2}{|c|}{$--$}& $p_{T}^{\ell\ell}>20$~GeV & $--$\\
\hline
\multirow{3}{*}{Jets} & \multicolumn{5}{|c}{$\vert y^j\vert < 4.4$, $\Delta R_{\ell,j}\ge 0.3$}\\
                      & \multicolumn{4}{|c|}{$p_{T}^{j_1}>55$~GeV} & $p_{T}^{j_1}>85$~GeV\\
                      & \multicolumn{4}{|c|}{$p_{T}^{j_2}>45$~GeV} & $p_{T}^{j_1}>75$~GeV\\
\hline
Dijet system          & $--$ & $m_{jj} > 1$~TeV & \multicolumn{2}{|c|}{$m_{jj} > 250$~GeV} & $--$\\
\hline
Interval jets         & \multicolumn{2}{|c|}{$--$} & $N_{\rm jet}^{\rm gap} = 0$ & $N_{\rm jet}^{\rm gap} \ge 1$ & $--$\\
\hline
Zjj system           & \multicolumn{2}{|c|}{$--$} & $p_{T}^{\rm balance} < 0.15$ & $p_{T}^{\rm balance,3} < 0.15$ & $--$\\
\hline
\hline
\end{tabular}
\caption{ Summary of the selection criteria for the $Z$jj measurement, that define the fiducial regions. ``Interval jets'' refer to
the selection criteria applied to the jets that lie in the rapidity interval bounded by the dijet system. The $p_T^{\rm balance}$ is defined as the ratio of the vectorial sum of leptons and two leading jets four momenta divided by their scalar sum; analogously the $p_T^{\rm balance,3}$ includes also the third leading jet in the sum.}
\label{tab:ZjjRegions}
\end{center}
\end{table}

In all five regions, the strong production dominates the expected composition, ranging from 85\% to 96\% of the total expectation, while electroweak contribution ranges between 1\% and 12\%. Other processes always contribute less than 4\% of the expected number of events.

\begin{figure}[htb]
\centering
\includegraphics[height=2in]{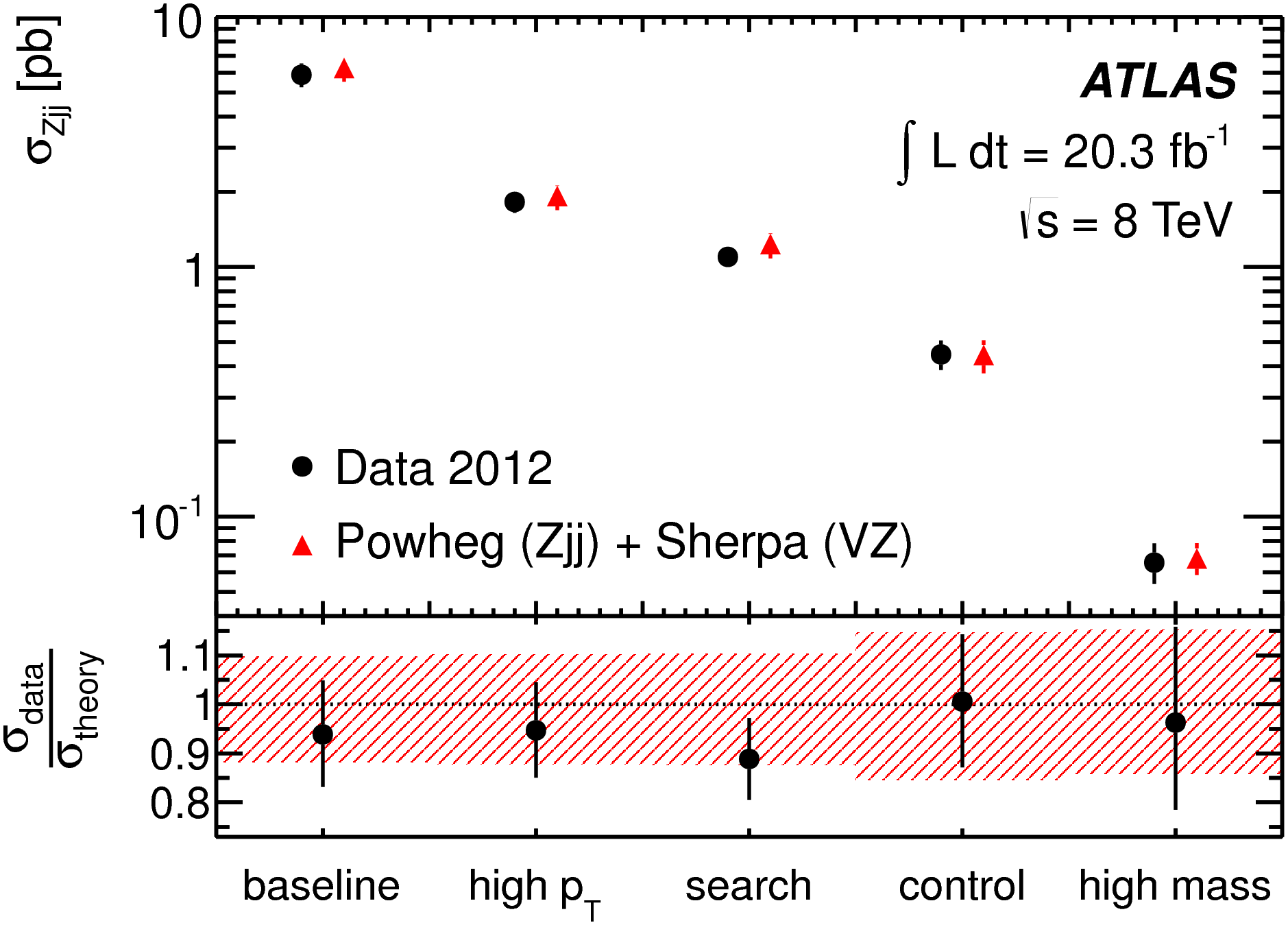}\hspace{1cm}
\includegraphics[height=2in]{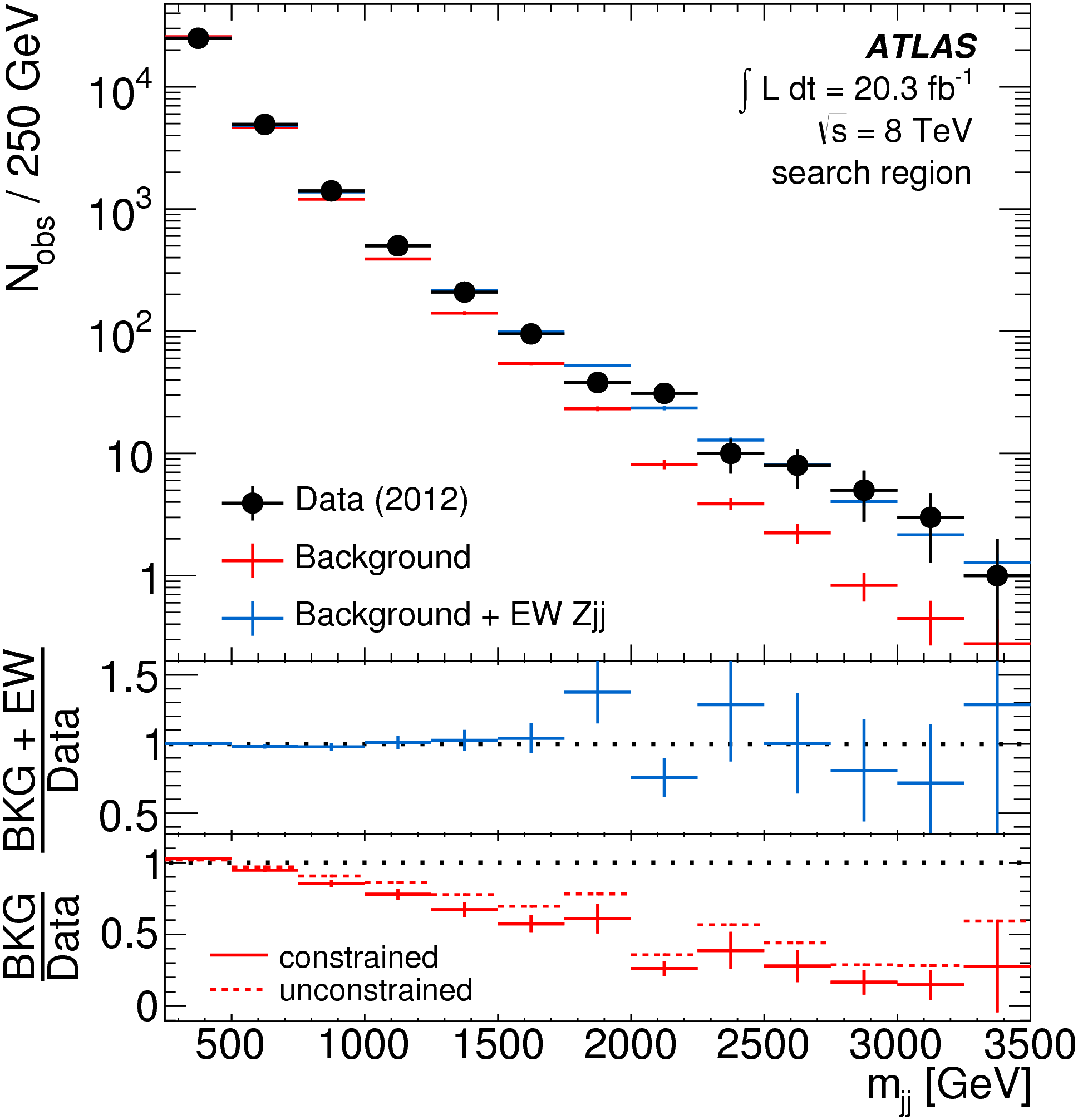}
\caption{Expected and measured fiducial cross sections for $Z$jj production. See text and Table~\ref{tab:ZjjRegions} for the definition of the fiducial regions. (right) Invariant mass of the two leading jets for the search region of the $Z$jj measurement. The histogram represents the result of the fit. The bottom panels show the ratio of the data and the signal+background or background-only prediction.}
\label{fig:Zjj_xsec}
\label{fig:Zjj_ewk}
\end{figure}

The fiducial cross section in each of the regions defined is measured as:
\begin{equation}
\sigma^{\rm fid.} = \frac{N({\rm data}) - N({\rm bkg.})}{L\cdot {\cal C}}
\end{equation}
\noindent
where N(data) is the number of observed events after selections, N(bkg) is the expected background, L is the integrated luminosity and {$\cal C$} is a correction factor estimated using simulation applying selections at reconstruction versus particle level. {\sc Sherpa}~\cite{sherpa} is used as generator for the simulation, and generated events are processed through a full ATLAS detector simulation~\cite{AtlasSimPaper}. The resulting fiducial cross sections are therefore defined at particle level, passing analogous selections to the ones presented in Table~\ref{tab:ZjjRegions}. The measured cross sections for the sum of electroweak and strong $Z$jj production for the five regions is shown in Figure~\ref{fig:Zjj_xsec}~(left). Good agreement with the theoretical expectation is observed in all five measurements within statistical and systematic uncertainties.

In addition, inclusive $Z$jj differential distributions are measured in all the five regions presented and corrected to particle-level. We refer to~\cite{Zjj} for a full list of results.

The extraction of the electroweak $Z$jj production component is performed in the {\it search} region, fitting the invariant mass of the two leading jets ($m(jj)$). The strong production is considered as a background. 
The signal template is obtained from a {\sc Sherpa} simulation of electroweak $Z$jj. The background
template is constructed from the {\sc Sherpa} strong Zjj simulation plus the small contribution
from other backgrounds. To reduce modeling uncertainties, the background $m(jj)$ distribution is constrained in the {\it control} region, exploiting the additional radiation expected in the case of strong production. A correction function is derived as function of $m(jj)$.

Figure~\ref{fig:Zjj_ewk}~(right) shows the $m(jj)$ distribution in the {\it search} region. The signal and background templates are normalized to the values obtained from the fit. A detector-corrected fiducial cross section for electroweak $Z$jj production is obtained:
\begin{equation}
\sigma^{\rm fid.}_{\rm ewk}(Zjj) = 54.7\pm4.6({\rm stat.})^{+9.8}_{-10.4}({\rm syst.})\pm 1.5({\rm lumi})~fb,
\end{equation}
\noindent
which is in agreement with the theoretical expectation obtained from Powheg~\cite{powhegbox}: $\sigma^{\rm fid, exp.}_{\rm ewk} = 46.1\pm 0.2)({\rm stat.})^{+0.3}_{-0.2}({\it scale})\pm 0.8({\rm PDF})\pm 0.5({\rm model})$~fb.

The dominant systematic uncertainties come from Jet Energy scale, from theory expectations and from the strong production modeling (transferring the re-weighting from the {\it control} to the {\it search} region. 
The background-only hypothesis is excluded at more than $5$ standard deviations, including both statistical and systematic uncertainties, using pseudo-experiments.

\section{Electroweak $W^{\pm}W^{\pm}$jj}

Single-lepton triggered events are used for the \wwjj~measurement. Two same-electric-charge leptons (either electrons or muons) are required to be have $p_T > 25~$GeV, $\vert\eta\vert < $2.5 and have a tight isolation requirement to suppress backgrounds producing non-prompt leptons. Events are divided into three channels depending on the lepton's flavor: $e^\pm e^\pm$, $e^\pm \mu^\pm$ and $\mu^\pm \mu^\pm$. Events are also required to have at least two jets with $p_T > 30$~GeV and $\vert\eta\vert < $4.5, where the minimum $p_T$ requirement is large enough to significantly reduce the sensitivity to pile-up effects.
A missing transverse energy of at least 40 GeV is required to exploit the expected neutrinos in the final state.
A tight veto on any additional identified electron or muon is used to reduce contribution from $WZ/\gamma^\star$ and $ZZ$ backgrounds; $WZ/\gamma^\star$ production, in association with two jets, still represent to largest background contribution to this analysis, accounting for up to $50\%$ of the expected background contribution and producing a comparable expected number of events to \wwjj~production.
 In addition, for the $e^\pm e^\pm$ channel only, the invariant mass of the two electrons must be outside a 10~GeV window around the $Z$ peak, to suppress background from Drell-Yan production where one electron charge is mis-reconstructed.
Events with at least one identified $b$-jet are vetoed, as well as events where an identified muon is close to a reconstructed jet ($\Delta R(\mu,jet)<0.3$), to reduce contribution from non-prompt backgrounds.

Two signal regions are defined. The first one is called {\it inclusive}, and is defined requiring, in addition to the above selection criteria, the invariant mass of the two leading $p_T$ jets to be above $500$~GeV. 
The fiducial cross section of both electroweak and strong \wwjj~production is measured in the {\it inclusive} region, considering both contribution as signal.
A subset of the {\it inclusive} region, obtained requiring the two leading jets to be separated in rapidity by $\vert\Delta y(jj)\vert > 2.4$, is labeled {\it VBS} region and the electroweak \wwjj~production cross section is measured after these selections. The strong \wwjj~production is therefore considered as a background in the {\it VBS} region.

The expected signal contribution is estimated using simulated events based on {\sc Sherpa} generator, with up to 3 jets in the Matrix Element, and full ATLAS simulation~\cite{AtlasSimPaper}. The same setup is used to estimate $WZ/\gamma^\star$ contribution. In both cases the expected fiducial cross section after particle-level selections for both the {\it inclusive} and {\it VBS} regions are normalized using state-of-the-art NLO calculation in QCD. The calculations are performed using a combination of Powheg~\cite{powhegbox,pb_ssww} and VBF@NLO~\cite{vbfnlo} generators, interfaced with Pythia~\cite{pythia,pythiaAtlasTunes} for parton-shower effects in the first case.
 Non-prompt leptons from charge mis-reconstruction arise mainly from asymmetric photon conversions. Contribution from $W\gamma$ is estimated using simulation (with the conversion rate checked in data), while contribution from Drell-Yan, $t\bar{t}$ and di-boson is estimated using a data-driven technique.
 Other non-prompt contributions are estimated using a fake-factor method based on events where one of the leptons is required to be non-isolated. Such contribution is dominated by $t\bar{t}$, $W+$jets and multi-jet production.

\begin{figure}[bht]
\centering
\includegraphics[height=2in]{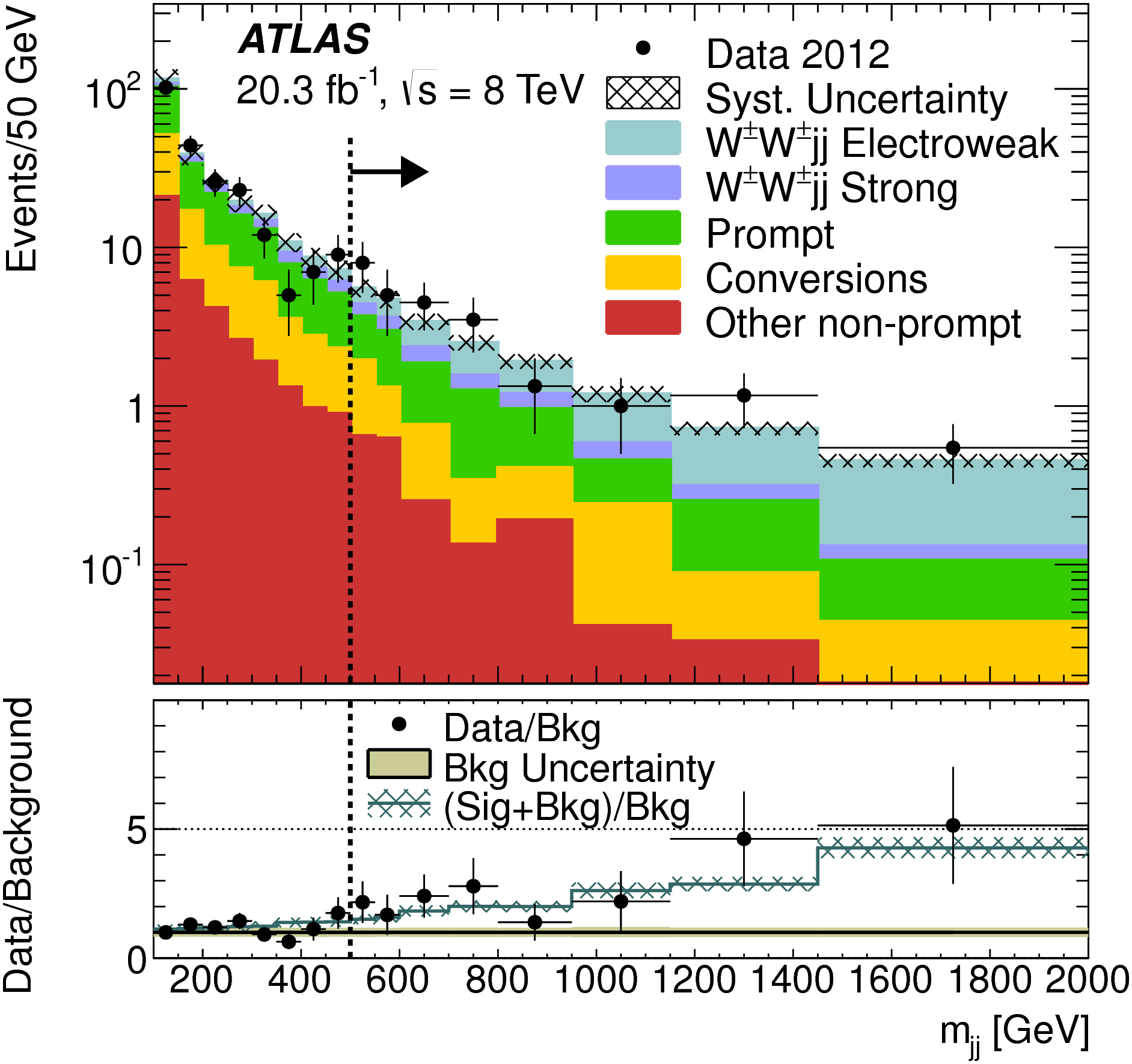}
\includegraphics[height=2in]{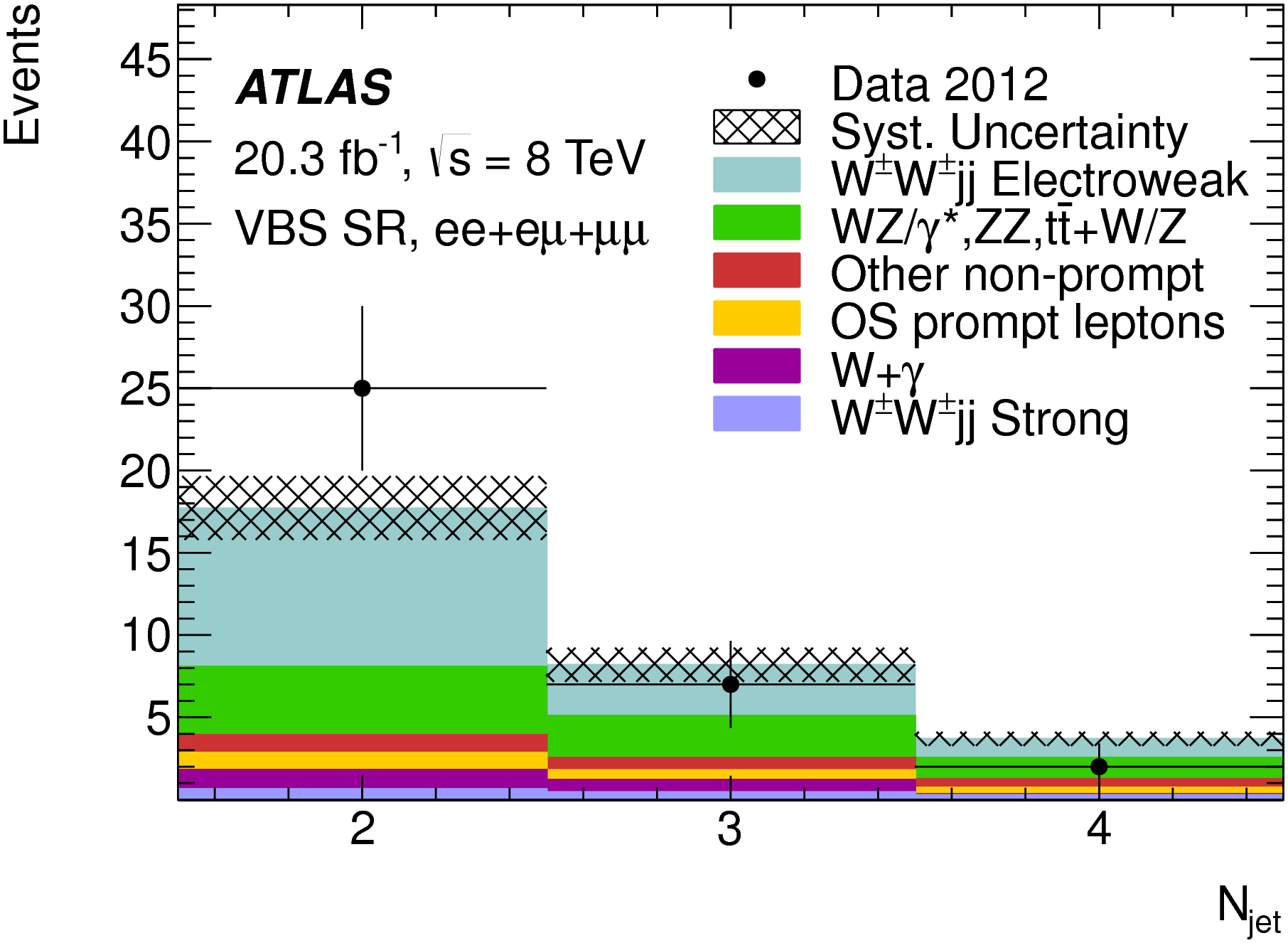}
\caption{(left) Invariant mass distribution of the two leading jets in the Inclusive signal region for the \wwjj~measurement. (right) Distribution of the number of jets in the VBS signal region for the \wwjj~measurement.}
\label{fig:WWjj_mjj}
\label{fig:WWjj_Njets}
\end{figure}

The fiducial cross section for electroweak and strong production is measured in the inclusive region, applying a correction for detector effects obtained from simulation. The resulting cross section, combining all three channels, is $\sigma^{\rm fid}_{inclusive}=2.1 \pm 0.5 ({\rm stat}) \pm 0.3 ({\rm syst})$~fb, in agreement with SM expectation of $1.52\pm0.11$~fb~\cite{WWjj,powhegbox,pb_ssww}. The background-only hypothesis is excluded at $4.5$ standard deviations, using pseudo-experiments. Figure~\ref{fig:WWjj_mjj}~(left) shows the distribution of expected and observed events as function of the invariant mass of the leading jets after {\it inclusive} region selections; the selection $m(jj)>500~$GeV defines the {\it inclusive} signal region.

The fiducial cross section for electroweak \wwjj~production is measured in the {\it VBS} signal region. The resulting fiducial cross section is $\sigma^{\rm fid}=1.3 \pm 0.4 ({\rm stat}) \pm 0.2 ({\rm syst})$~fb, in agreement with the SM expectation ($0.95\pm 0.06$~fb)~\cite{WWjj,powhegbox,pb_ssww}. For this measurement, the strong production is considered as a background and background-only hypothesis is excluded at 3.6 standard deviations.

A particularly interesting kinematic distribution in the signal region is the number of reconstructed jets, shown in Figure~\ref{fig:WWjj_Njets}~(right). In the case of electroweak production the jet multiplicity above two is expected to drop rapidly, while this is not the case for strong production. Data show the same drop observed in simulation, nicely confirming the expected behavior.

\section{Conclusions}
This contribution has presented the first measurements of electroweak $Z$ and $W^\pm W^\pm$ production in association with two jets. Fiducial cross sections have been measured for both strong and electroweak production mechanisms together, as well as for the electroweak production mechanism alone. Sensitive kinematic distributions have been unfolded at particle-level for the $Z$jj final state, while for \wwjj~the statistical error is still large enough to not fully justify such approach at this time. 

During the next LHC data-taking run, we expect a larger production cross section (about a factor of 4), due to the higher center of mass energy, and more data to be collected in the next years of operations. The presented measurements pave the road for a deeper investigation of the electroweak symmetry breaking using vector-boson-fusion and vector-boson-scattering topologies.

\end{document}